%% file: root.tex
\documentclass[a4paper, 10pt, conference]{ieeeconf} 

\IEEEoverridecommandlockouts                              
                                                         
\usepackage{xcolor}
\usepackage{cite}
\usepackage{amsmath,amssymb,amsfonts}
\usepackage{algorithmic}
\usepackage{algorithm}
\usepackage{graphicx}
\usepackage{subfig}
\usepackage{textcomp}

\graphicspath{{pics/}}

\input{./include/pub_tools.tex}

\input{./include/glossaries.tex}

\title{\LARGE \bf
Grid-Based Stochastic Model Predictive Control for Trajectory Planning in Uncertain Environments
}

\author{Tim~Br\"udigam$^{1}$,~Fulvio~Di~Luzio$^{2}$,~Lucia~Pallottino$^{2}$,~Dirk~Wollherr$^{1}$,~and~Marion~Leibold$^{1}$
\thanks{*The authors gratefully acknowledge the financial 
support by the BMW Group within the CAR@TUM project.}
\thanks{$^{1}$T. Br\"udigam, D. Wollherr, and M. Leibold are with the Chair of Automatic Control Engineering at the Technical University of Munich, Germany.
{\tt\small \{tim.bruedigam;~dw;~marion.leibold\}@tum.de}}%
\thanks{$^{2}$F. Di Luzio and L. Pallottino are with the Department of Information Engineering at the University of Pisa, Italy. L. Pallottino is also with the Research Center E. Piaggio of the University of Pisa.
        {\tt\small f.diluzio@studenti.unipi.it; lucia.pallottino@unipi.it}}%
}

\begin{document}

\maketitle
\thispagestyle{empty}
\pagestyle{empty}


\begin{abstract}
Stochastic Model Predictive Control has proved to be an efficient method to plan trajectories in uncertain environments, e.g., for autonomous vehicles. Chance constraints ensure that the probability of collision is bounded by a predefined risk parameter. However, considering chance constraints in an optimization problem can be challenging and computationally demanding. In this paper, we present a grid-based Stochastic Model Predictive Control approach. This approach allows to determine a simple deterministic reformulation of the chance constraints and reduces the computational effort, while considering the stochastic nature of the environment. Within the proposed method, we first divide the environment into a grid and, for each predicted step, assign each cell a probability value, which represents the probability that this cell will be occupied by surrounding vehicles. Then, the probabilistic grid is transformed into a binary grid of admissible and inadmissible cells by applying a threshold, representing a risk parameter. Only cells with an occupancy probability lower than the threshold are admissible for the controlled vehicle. Given the admissible cells, a convex hull is generated, which can then be used for trajectory planning. Simulations of an autonomous driving highway scenario show the benefits of the proposed grid-based Stochastic Model Predictive Control method.
\end{abstract}

\input{./include/ieee_header3.tex}
\input{./chapters/introduction.tex}

\input{./chapters/problem.tex}

\input{./chapters/method.tex}

\input{./chapters/results.tex}

\input{./chapters/discussion.tex}

\input{./chapters/conclusion.tex}


%



%





\bibliography{./references/refs_fulvio,./references/Dissertation_bib}
\bibliographystyle{unsrt}

\input{./include/ieee_header4.tex}
\end{document}

%% file: include/pub_tools.tex
\usepackage{mathtools}

\usepackage[acronym, style=alttree, toc=true, shortcuts, xindy, nomain, nonumberlist]{glossaries}

\usepackage{cleveref}

\usepackage{hyphenat}
\usepackage{bm}

\usepackage{siunitx}  

\newcommand{\norm}[1]{\left\lVert#1\right\rVert} 

\newcommand{\lr}[1]{\left(#1\right)} 

\DeclarePairedDelimiterXPP\onenorm[1]{}\lVert\rVert{_1}{\ifblank{#1}{\:\cdot\:}{#1}} 
\DeclarePairedDelimiterXPP\twonorm[1]{}\lVert\rVert{_2}{\ifblank{#1}{\:\cdot\:}{#1}} 

\newglossary{symbols}{sym}{sbl}{List of Symbols}
\newglossary{notation}{not}{nt}{Notation}

\glsaddkey{dimension}{\glsentrytext{\glslabel}}{\glsentryd}{\GLsentryd}{\glsd}{\Glsd}{\GLSd}
\glsaddkey{body}{\glsentrytext{\glslabel}}{\glsentrybody}{\GLsentrybody}{\glsbody}{\Glsbody}{\GLSbody}

%% file: include/glossaries.tex
\newacronym{MPC}{MPC}{Model Predictive Control}
\newacronym{MPCa}{MPC algorithm}{MPC algorithm}
\newacronym{RMPC}{RMPC}{Robust Model Predictive Control}
\newacronym{SMPC}{SMPC}{Stochastic Model Predictive Control}
\newacronym{SCMPC}{SCMPC}{Scenario Model Predictive Control}
\newacronym{MILP}{MILP}{Mixed Integer Linear Program}
\newacronym{PIT}{PIT}{Pointwise\hyp{}In\hyp{}Time}
\newacronym{POMDP}{POMDP}{Partially Observable Markov Decision Process}
\newacronym{MDP}{MDP}{Markov Decision Process}
\newacronym{KKT}{KKT}{Karush\hyp{}Kuhn\hyp{}Tucker}
\newacronym{ev}{EV}{ego vehicle}
\newacronym{tv}{TV}{target vehicle}
\newacronym{cog}{CoG}{center of gravity}
\newacronym{ol}{OL}{Open-Loop}
\newacronym{cl}{CL}{Closed-Loop}
\newacronym{ocp}{OCP}{Optimal Control Problem}
\newacronym{pog}{PG}{\textit{Probabilistic Grid}}
\newacronym{bog}{BG}{\textit{Binary Grid}}
\newacronym{lk}{LK}{lane keeping}
\newacronym{lc}{LC}{lane changing}
\newacronym{og}{OG}{Occupancy Grid}

	\newglossaryentry{lr}{type=symbols,
		sort={vehicle},
		name={\ensuremath{l_\text{r}}},
		description={...}
	}
	
	\newglossaryentry{lf}{type=symbols,
	sort={vehicle},
	name={\ensuremath{l_\text{f}}},
	description={...}
	}

	\newglossaryentry{xitv}{type=symbols,
	sort={vehicle},
	name={\ensuremath{\boldsymbol{\xi}^\text{TV}}},
	description={...}
	}

	\newglossaryentry{xiev}{type=symbols,
		sort={vehicle},
		name={\ensuremath{\boldsymbol{\xi}^\text{EV}}},
		description={...}
	}

	\newglossaryentry{df}{type=symbols,
		sort={vehicle},
		name={\ensuremath{\delta_\text{f}}},
		description={...}
	}

	\newglossaryentry{cx}{type=symbols,
		sort={vehicle},
		name={\ensuremath{l_x}},
		description={...}
	}

	\newglossaryentry{cy}{type=symbols,
		sort={vehicle},
		name={\ensuremath{l_y}},
		description={...}
	}
	
	\newglossaryentry{ctv}{type=symbols,
		sort={vehicle},
		name={\ensuremath{\bm{c}^\text{TV}_{h}}},
		description={...}
	}
	
	\newglossaryentry{pth}{type=symbols,
	sort={vehicle},
	name={\ensuremath{p_\text{th}}},
	description={...}
	}
	
	\newglossaryentry{rl}{type=symbols,
		sort={road},
		name={\ensuremath{r_l}},
		description={road length}
	}
	
	\newglossaryentry{rw}{type=symbols,
		sort={road},
		name={\ensuremath{r_w}},
		description={road width}
	}

	\newglossaryentry{lref}{type=symbols,
		sort={vehicle},
		name={\ensuremath{l_\text{ref}}},
		description={reference lane}
	}
	
	\newglossaryentry{tv1}{type=symbols,
		sort={vehicle},
		name={\ensuremath{\text{TV}_1}},
		description={reference lane}
	}

	\newglossaryentry{tv2}{type=symbols,
		sort={vehicle},
		name={\ensuremath{\text{TV}_2}},
		description={reference lane}
	}

	\newglossaryentry{xitv1}{type=symbols,
		sort={vehicle},
		name={\ensuremath{\boldsymbol{\xi}^{\text{TV}_1}}},
		description={...}
	}

	\newglossaryentry{xitv2}{type=symbols,
		sort={vehicle},
		name={\ensuremath{\boldsymbol{\xi}^{\text{TV}_2}}},
		description={...}
	}


%% file: chapters/introduction.tex
\section{Introduction}
\label{sec:introduction}

\vspace{-15cm}
\mbox{\small 
This~work~has~been~accepted~to~the~IEEE~2020~International~Conference~on~Intelligent~Transportation~Systems.}
\vspace{14.2cm}

\vspace{-14.6cm}
\mbox{\small The~published~version~may~be~found~at~https://doi.org/10.1109/ITSC45102.2020.9294388.}
\vspace{13.6cm}

\gls{MPC} has shown to be an efficient method to control autonomous systems, especially autonomous vehicles \cite{LevinsonEtalThrun2011, 
KatrakazasEtalDeka2015, 
GutjahrGroellWerling2017}. \gls{MPC} is an optimization-based control method which iteratively solves an optimization problem, possibly with constraints, on a finite horizon. The benefit of iteratively solving an optimization problem with hard constraints enables planning trajectories and maneuvers in dynamic environments without collisions. A major challenge for autonomous driving is to handle uncertainties, especially future behavior of other traffic participants. Exact maneuver predictions are impossible and modeling the exact execution of maneuvers is not perfectly accurate. Therefore, uncertainties must be included in the \gls{MPC} problem formulation.

While \gls{RMPC} yields robust and safe solutions, these solutions are often impractical in dense traffic or for longer prediction horizons, as accounting for the worst-case uncertainty realizations results in overly conservative trajectory planning. Conservatism is reduced by applying \gls{SMPC} \cite{Mesbah2016} or \gls{SCMPC}~\cite{SchildbachEtalMorari2014} with chance constraints. In contrast to robust hard constraints, a chance constraint is only required to be satisfied according to a predefined risk parameter, allowing less conservatively planned trajectories, as worst-case uncertainty realizations are not considered. Combining \gls{RMPC} and \gls{SMPC} for mobile robots is considered in \cite{BruedigamEtalLeibold2020b}.

\gls{SMPC} \cite{CarvalhoEtalBorrelli2014, LenzEtalKnoll2015} and \gls{SCMPC} \cite{SchildbachBorrelli2016, CesariEtalBorrelli2017} have previously been applied in autonomous driving. A combination of \gls{SMPC} and \gls{SCMPC} to handle uncertainty of surrounding vehicles is suggested in \cite{BruedigamEtalWollherr2018b} where \gls{SCMPC} accounts for maneuver uncertainty and \gls{SMPC} copes with maneuver execution uncertainty of other vehicles. However, solving the optimization problem with chance constraints often requires Gaussian probability distributions or considers only the most likely future motion of other vehicles to simplify the predictions. Additionally, considering an increased number of dynamic objects with chance constraints can be computationally challenging.

An \gls{og} \cite{thrun2005} is a mapping grid of the environment where each grid cell is assigned a probability that a certain area is occupied. In \cite{cue2} and \cite{tay} first approaches of \glspl{og} for autonomous vehicles have been proposed. In \glspl{og} concepts like objects, pedestrians, and vehicles do not exist and data fusion from multiple sensors is efficient. As summarized in \cite{bofreview}, \glspl{og} have been developed in many ways with focus on autonomous driving for both highway and urban traffic scenarios, e.g., in \cite{SteyerTanzmeisterWollherr2018, SteyerEtalWollherr2019}. However, research has been mainly carried out on how to treat data in order to determine a correspondence between sensors and grid cells, and how occupancy probability is assigned and updated with the accumulation of new data.%

In this paper we present a grid-based \gls{SMPC} method for trajectory planning in uncertain environments. While it is possible to apply the proposed method to various autonomous systems, here we will focus on autonomous vehicles. The proposed method provides a simple strategy to handle arbitrary uncertainty of future vehicle motion. The computational effort of the \gls{SMPC} optimization problem is manageable and the approach scales well with an increased number of surrounding vehicles or other obstacles. We consider a grid for the environment, i.e., the road. For every predicted step, each cell then gets assigned a probability value, representing its occupancy probability by an obstacle. All cells with an occupancy probability value larger than a predefined \gls{SMPC} risk parameter are inadmissible, where the risk parameter works as a threshold. Given the admissible cells, convex admissible state constraints are defined for the optimization problem. 

The two main benefits of using a grid-based \gls{SMPC} approach for autonomous driving, compared to other \gls{SMPC} approaches, are the following. 
First, it is not required to generate and consider an individual chance constraint for each vehicle or obstacle considered. The probability grid is generated given all obstacles and then the risk parameter threshold is applied to all cells, yielding a deterministic reformulation of the chance constrained optimization problem. This results in an optimization problem with convex state constraints, which can be solved efficiently. However, the stochastic nature of the problem is still accounted for as a probabilistic grid is initially generated. Second, it is not necessary to decide on a most likely behavior of other vehicles, as multiple predicted behavior options with arbitrary probability distribution can be considered. These properties facilitate the application to autonomous driving. The effectiveness of the presented approach is demonstrated in a highway simulation. 

The remainder of the paper is structured as follows. Section \ref{sec:problem} introduces the vehicle models and in Section \ref{sec:method} the method is derived in detail. Section \ref{sec:results} shows a highway simulation with the proposed method, followed by conclusive remarks in Section \ref{sec:conclusion}.

%% file: chapters/problem.tex
\section{Vehicle Models}
\label{sec:problem}

\gls{MPC} requires system models for the prediction of futures states within the prediction horizon. We specifically consider trajectory planning for vehicles, where the controlled vehicle is known as the \gls{ev} and surrounding vehicles as \glspl{tv}. Focusing on vehicles allows to present the proposed gird-based \gls{SMPC} method more comprehensively. However, \glspl{tv} can be interpreted as dynamic obstacles in non-vehicle related trajectory planning tasks.

We consider a nonlinear \gls{ev} model 
\begin{IEEEeqnarray}{c}
	\IEEEyesnumber \label{eq:systemEV_general}
	\dot{\boldsymbol{\xi}}^\text{EV} = f \lr{\boldsymbol{\xi}^\text{EV}, \boldsymbol{u}^\text{EV}}
\end{IEEEeqnarray}
with \gls{ev} state $\boldsymbol{\xi}^\text{EV}$ and \gls{ev} input $\boldsymbol{u}^\text{EV}$. For the discrete optimization problem model \eqref{eq:systemEV_general} needs to be discretized. Control constraints are imposed on both the steering angle and the acceleration, i.e., $\bm{u}_\text{min} \leq \bm{u} \leq \bm{u}_\text{max}$, 
summarized as $\bm{u} \in \mathcal{U}$. The \gls{ev} is subject to state constraints $\boldsymbol{\xi}^\text{EV} \in \Xi$, such as road restrictions, and specifically safety constraints $\boldsymbol{\xi}^\text{EV} \in \Xi^\text{safe}$, which ensure collision avoidance with other vehicles.

It is necessary for the \gls{ev} to predict the future \gls{tv} motion. It is assumed that the future \gls{tv} motion is described by a linear discrete-time point-mass model subject to prediction uncertainty, similar to \cite{BruedigamEtalWollherr2018b}, 
\begin{IEEEeqnarray}{rl}
	\IEEEyesnumber \label{eq:systemTV}
		\gls{xitv}_{k+1} &= \boldsymbol{A} \gls{xitv}_k + \boldsymbol{B}\bm{u}_k^\text{TV} + \boldsymbol{G} \bm{w}_k^\text{TV}
\end{IEEEeqnarray}
where $\boldsymbol{\xi}_k^\text{TV} = \left[ x_k^\text{TV},v_{x,k}^\text{TV},y_k^\text{TV},v_{y,k}^\text{TV} \right]^\top$ is the TV state at time step $k$, represented by longitudinal and lateral positions and velocities, and $\boldsymbol{u}_k^\text{TV} = \left[ u_{x,k}^\text{TV},u_{y,k}^\text{TV} \right]^\top$ is the control input consisting of longitudinal and lateral acceleration with the assumed to be known \gls{tv} reference trajectory $\gls{xitv}_{k,\text{ref}}$ and feedback law
\begin{IEEEeqnarray}{c}
	\label{eq:uTV} 
	\boldsymbol{u}_k^\text{TV} = \boldsymbol{K} \lr{\gls{xitv}_k - \gls{xitv}_{k,\text{ref}}},~~\boldsymbol{K} = \begin{bmatrix}
						0 & k_{12} & 0 & 0 \\
						0 & 0 & k_{21} & k_{22}
					  \end{bmatrix}. \IEEEeqnarraynumspace
\end{IEEEeqnarray}
\noindent
The uncertainty in the prediction is taken into account by the random variable $\boldsymbol{w}_k^\text{TV}$ and $\boldsymbol{G} = \text{diag}\left(g_1,g_2,g_3,g_4\right)$. 

The presented vehicle models are then used in the \gls{MPC} optimization problem to predict the future \gls{ev} and \gls{tv} states.

%% file: chapters/method.tex
\section{Method}
\label{sec:method}


This section presents the main contribution of this work. Surrounding \glspl{tv} have uncertain behavior. EV safety constraints must therefore consider the stochastic nature of the future \gls{tv} motion to avoid collisions. Robustly accounting for the worst-case uncertainty results in overly conservative vehicle trajectories. Introducing chance constraints in \gls{SMPC} allows to relax this conservatism, depending on a tunable risk parameter. The probabilistic chance constraints need to be reformulated into deterministic expressions, so that they can be solved within the optimization problem. 

Assuming a grid representation of the environment, we derive a grid-based \gls{SMPC} approach, which enables a simple approach to reformulate the probabilistic chance constraints into tractable constraints. First, for each time step of the \gls{SMPC} prediction horizon a probabilistic occupancy grid is computed, where each cell of the grid represents the probability that the cell is occupied by a \gls{tv}. This leads to the formulation of chance constraints to avoid collisions between the \gls{ev} and \gls{tv}. A tractable expression of the chance constraint is found by deriving a binary grid in order to clearly identify the admissible road grid cells and by finding a convex hull in which the \gls{ev} can operate. Finally, we solve the optimal control problem of the \gls{SMPC}. In the following subsections we describe the method in detail, starting with the probabilistic occupancy grid. Explicitly denoting the current time step $k$ for states is omitted due to clarity. Prediction steps are indicated by the index $h$.

\subsection{Probabilistic Grid}

The environment is represented by a grid $\mathcal{G}$, i.e., an evenly spaced field of cells $ \bm{c}_{i,j}  \in \mathcal{G}$ with $\bm{c}_{i,j}  = (c_i, c_j)$. Each cell has dimensions \gls{cx} and \gls{cy} accounting for length and width, respectively, and is identified in a 2D space by two indices $i$ and $j$. The level of approximation depends on the size of the cell.

For every prediction step $h$ an individual grid is generated, which is later used for collision avoidance. Here, the \gls{pog} is inspired by \glspl{og} but it is defined differently compared to standard \gls{og} literature. The \gls{pog}, represented by a matrix $\mathcal{P}$, consists of elements $p_{i,j}$ which describe the occupancy probability. However, the probability value $p_{i,j}$ does not necessarily correspond to the exact probability that cell $\bm{c}_{i,j}$ is occupied by a TV. This is necessary as the proposed method later combines grids of individual \glspl{tv}. 

For simplicity, in the following we assume a Gaussian probability distribution for the \gls{tv} motion prediction in order to demonstrate the method. However, it is possible to apply arbitrary probability distributions within the proposed method. The probability distribution, in this case a 2D Gaussian distribution, is used to assign probability values to the matrix $\mathcal{P}$. The probability density function is given by
\begin{IEEEeqnarray}{c}
	f_{c^\text{TV}} \left( \bm{c} \right) = \frac{\mathrm{exp}\lr{-\frac{1}{2}\lr{\bm{c} -\bm{c}^\text{TV}_{h}}^\top \boldsymbol{\Sigma_}{h}^{-1}\lr{\bm{c} -\bm{c}^\text{TV}_{h}}}}{\sqrt{(2\pi)^2 |\boldsymbol{\Sigma}_{h}|}} 
 \label{eq:2DGauss} \IEEEeqnarraynumspace
\end{IEEEeqnarray}
\noindent
where $\bm{c}^\text{TV}_{h}$ is the cell corresponding to the estimated TV center at prediction step $h$ and $\Sigma_{h}$ is a covariance matrix. 

The uncertainty $\boldsymbol{\Sigma}_{h}$ in \eqref{eq:2DGauss} is propagated by using a recursive technique, described in \cite{CarvalhoEtalBorrelli2014}, \begin{equation}\label{varianceprop}
\boldsymbol{\Sigma}_{h+1} = \left( \boldsymbol{A}+\boldsymbol{B}\boldsymbol{K} \right)\boldsymbol{\Sigma}_{h}\left( \boldsymbol{A}+\boldsymbol{B}\boldsymbol{K} \right)^\top + \boldsymbol{G} \boldsymbol{\Sigma}_w \boldsymbol{G}^\top
\end{equation} 
given the assumed \gls{tv} prediction model \eqref{eq:systemTV}, the initial condition $\boldsymbol{\Sigma}_{0} = \boldsymbol{0}$, and the uncertainty covariance matrix~$\boldsymbol{\Sigma}_w$.

\begin{figure}[!htbp]
	\centering
	\subfloat[Probability occupancy grid given \gls{tv} \gls{cog}. \label{bell:1}]{\includegraphics[scale=0.3,trim= 0 0 0 0]{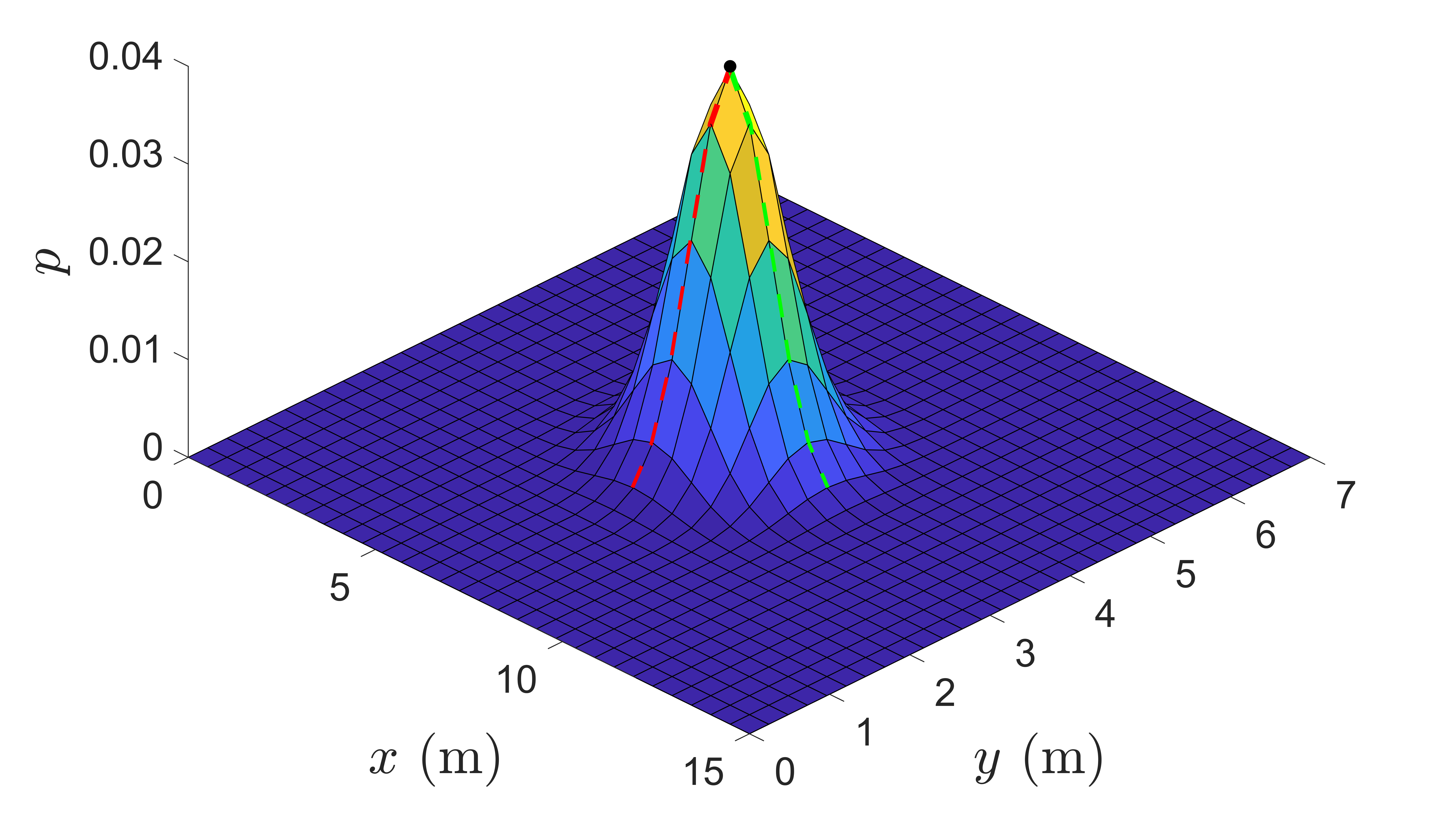}}
	\\
	\subfloat[Probability occupancy grid taking into account the \gls{tv}
	dimensions. \label{bell:3}]{\includegraphics[scale=0.3,trim= 0 0 0 0]{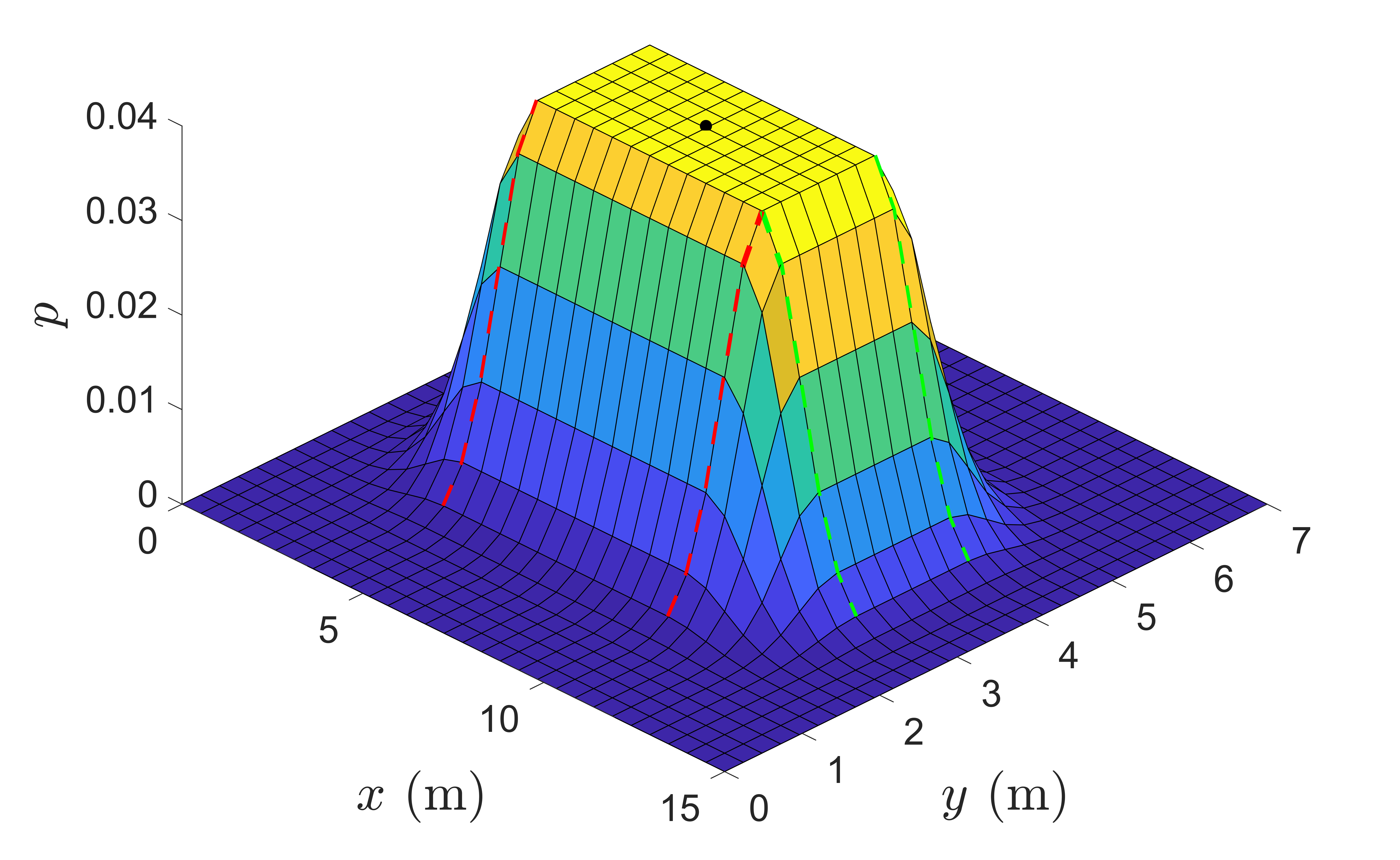}}
	\caption{Illustration for building the \gls{tv} probability occupancy grid. The black point represents \gls{ctv}.}
	\label{bell:all}
\end{figure}

In order to consider the \gls{tv} dimensions, we first compute probability values for each cell according to \eqref{eq:2DGauss}, displayed in Figure~\ref{bell:1}. The black point represents~\gls{ctv}, i.e., the \gls{cog} of the \gls{tv}, the green dashed line represents the normal distribution on the $xp$-plane passing through \gls{ctv}, while the red dashed line represents the normal distribution on the $yp$-plane. Then, we start to expand the maximum value computed with \eqref{eq:2DGauss} along both dimensions $x$ and $y$ in order to cover the vehicle area. 
The final result is displayed in Figure~\ref{bell:3}, where the distance between the two green dashed lines equals the width of the vehicle, and the distance between the red dashed lines equals the length. Note that Figure~\ref{bell:3} does not show a probability density function, as the maximal value was artificially expanded.  

In the following, the \gls{pog} is extended for the case of multiple vehicles.

\subsection{Multiple Target Vehicles and Maneuver Probabilities} \label{sec:scenarioext}
The presented \gls{pog} can easily be extended to scenarios with more than one \gls{tv} by computing a \gls{pog} for each TV and then adding up the probability values of the different grids for each cell. If a certain area on the road can potentially be occupied by more than one vehicle in the future, its probability to be occupied increases. This yields
\begin{IEEEeqnarray}{rl}
	\IEEEyesnumber \label{eq:ext1}
	p_{i,j} &= \sum_{n_\text{v}=1}^{N_\text{v}} p_{i,j,n_\text{v}}
\end{IEEEeqnarray}
\noindent
where $N_\text{v}$ is the number of \glspl{tv} localized in the detection range of the \gls{ev}.

\begin{figure}
    \centering
    \includegraphics[width = \columnwidth]{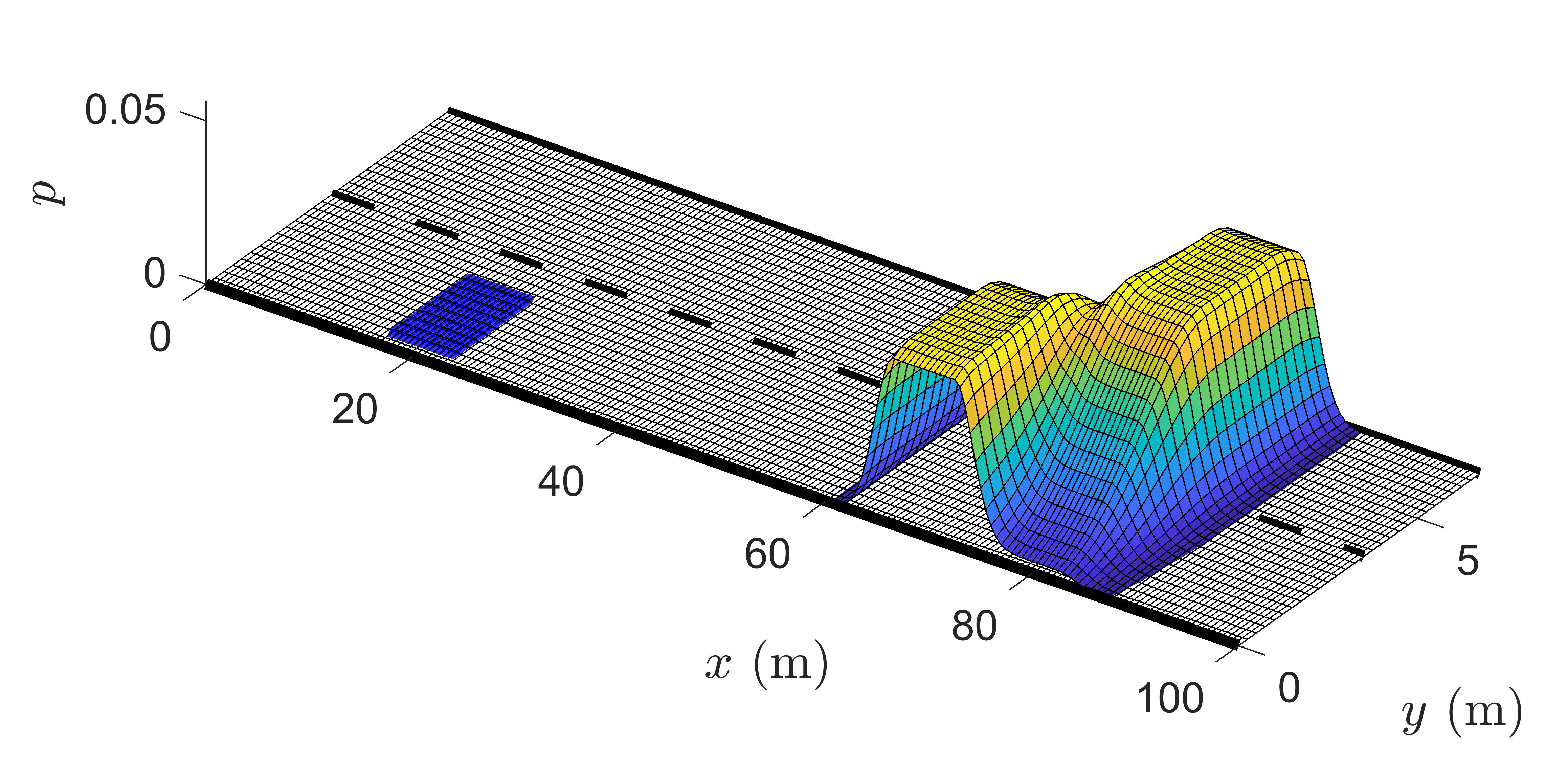}
    \caption{Probabilistic Grid of a prediction step. The \gls{ev} is on the left in blue and two TVs are displayed on the right.}
    \label{fig:combinedgrid}
\end{figure}

We additionally consider $N_\text{m}$ possible \gls{tv} maneuvers. This is achieved by computing a \gls{pog} for each maneuver and weigh it with the probability that this maneuver is actually performed, resulting in
\begin{IEEEeqnarray}{rl}
	\IEEEyesnumber \label{eq:ext2}
	p_{i,j} &= \sum_{n_\text{v}=1}^{N_\text{v}} \sum_{n_\text{m}=1}^{N_\text{m}} \varepsilon_{n_\text{v},n_\text{m}} p_{i,j,n_\text{v}, n_\text{m}} 
\end{IEEEeqnarray}
\noindent
where $\varepsilon_{n_\text{v},n_\text{m}}$ is the probability that the maneuver $n_\text{m}$ is performed by \gls{tv} $n_\text{v}$. A resulting example grid with two \glspl{tv} is shown in Figure~\ref{fig:combinedgrid}.

In the following, the \gls{pog}, considering all surrounding \glspl{tv}, is adapted to be included in an \gls{SMPC} optimization problem.

\subsection{Binary Grid} \label{sec:methodc}

In \gls{SMPC} the probabilistic chance constraint is required to be reformulated into a deterministic expression. We consider the chance constraint
\begin{IEEEeqnarray}{rl}
	\IEEEyesnumber \label{eq:chance}
\text{Pr}\left( \boldsymbol{\xi}^\text{EV} \in \Xi^\text{safe} \right) \geq \beta
\end{IEEEeqnarray}
where $\Xi^\text{safe}$ is the set of state which guarantees collision avoidance. The safety constraint $\boldsymbol{\xi}^\text{EV} \in \Xi^\text{safe}$ is required to be satisfied with probability $\beta$, where $\beta$ is a tunable risk parameter. Calculating a tractable, deterministic representation of the chance constraint can be computationally expensive and also challenging for non-Gaussian probability distributions.

To tackle this problem, we reformulate the chance constraints for the \gls{SMPC} by utilizing the presented \gls{pog}. We transform the probabilistic grid $\mathcal{P}$ into a \gls{bog}, represented by a matrix $\mathcal{B}$ with elements $b_{i,j}$, by imposing a probability threshold \gls{pth} similar to the risk parameter $\beta$,
\begin{IEEEeqnarray}{rl}
	\IEEEyesnumber \label{eq:bog}
	b_{i,j} &= \left \{ 
	\begin{aligned}
		&1, && \text{if}\ p_{i,j} \geq \gls{pth},\\
		&0, && \text{otherwise.}\
	\end{aligned} \right.
\end{IEEEeqnarray}
\noindent
The value $1$ indicates that a certain cell $\bm{c}_{i,j}$ is considered to be occupied and therefore inadmissible. The parameter \gls{pth} is a trade-off between risk and conservatism: the lower the threshold, the more conservative the controller. By transforming the probabilistic grid $\mathcal{P}$ into the binary grid~$\mathcal{B}$, we make a clear distinction between admissible and inadmissible space for the \gls{ev}. 

Now, the chance constraint in \eqref{eq:chance} can be expressed as a standard hard constraint
\begin{IEEEeqnarray}{rl}
	\IEEEyesnumber \label{eq:det}  
\boldsymbol{\xi}^\text{EV} \in \Xi^\text{adm}
\end{IEEEeqnarray}
\noindent
where $\Xi^\text{adm}$ includes all admissible cells in the \gls{bog}. The constraint \eqref{eq:det} can be handled in a general optimization problem. 

This approach resembles the use of chance constraints in other \gls{SMPC} approaches, as cells with low occupancy probability are neglected, depending on the threshold \gls{pth}. We then obtain constraints \eqref{eq:det} that can be directly handled by a solver. In the next subsection we show how to derive a linear inequality description of the constraint in \eqref{eq:det}. 

\subsection{Convex Admissible Safe State Space}
In order to obtain a fast \gls{SMPC} framework, it is not sufficient to define a deterministic constraint reformulation, as seen above, it is also beneficial to find a linear inequality description of the safety constraint. This is achieved by applying Bresenham's line algorithm \cite{Bresenham1965AlgorithmFC}. 

Given two points in a grid, which are connected by a straight line, Bresenham's line algorithm yields all cells which are touched by the connecting line.  
Even if it was developed in the field of computer graphics to select the bitmaps of an image, it is applicable here - instead of bitmaps we have grid cells. 

The goal here is to find a convex hull for all admissible cells of the \gls{bog}, applying Bresenham's algorithm. The convex hull, consisting of valid cells, can be described by linear constraints. 

\begin{figure}[!htbp]
	\centering
	\subfloat[The yellow cell represents the \gls{ev} center $\boldsymbol{c}^\text{EV}$, brown cells show the rear corners of the \gls{ev}, cyan cells ($\boldsymbol{C}_\text{range}$) indicate the cells in the (approximated) \gls{ev} detection range. Bresenham's algorithm evaluates if cells are free between the rear corner cells and the detection range cells (orange line). If a free path from a cell $\boldsymbol{c}$ of $\boldsymbol{C}_\text{range}$ to both $c_{\text{r},1}$ and $c_{\text{r},2}$ is verified, the cell is marked with a green color ($\boldsymbol{C}_\text{free}$), otherwise with red.
	\label{hull:1}]{\includegraphics[scale=0.20]{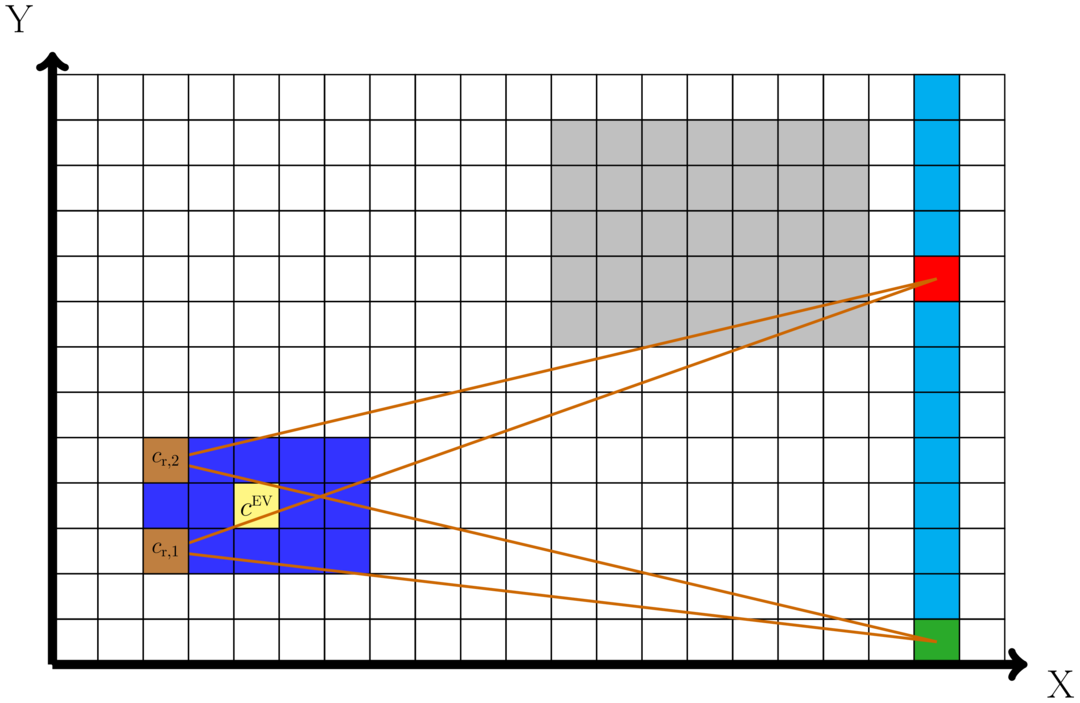}}
	\\
	\subfloat[The edges of the green column are marked as $e_1$ and $e_2$. To obtain a larger convex hull, the rear corner cells are moved outside the vehicle bounds if connections are possible to $e_1$ and $e_2$ without intersecting inadmissible cells. \label{hull:2}]{\includegraphics[scale=0.20]{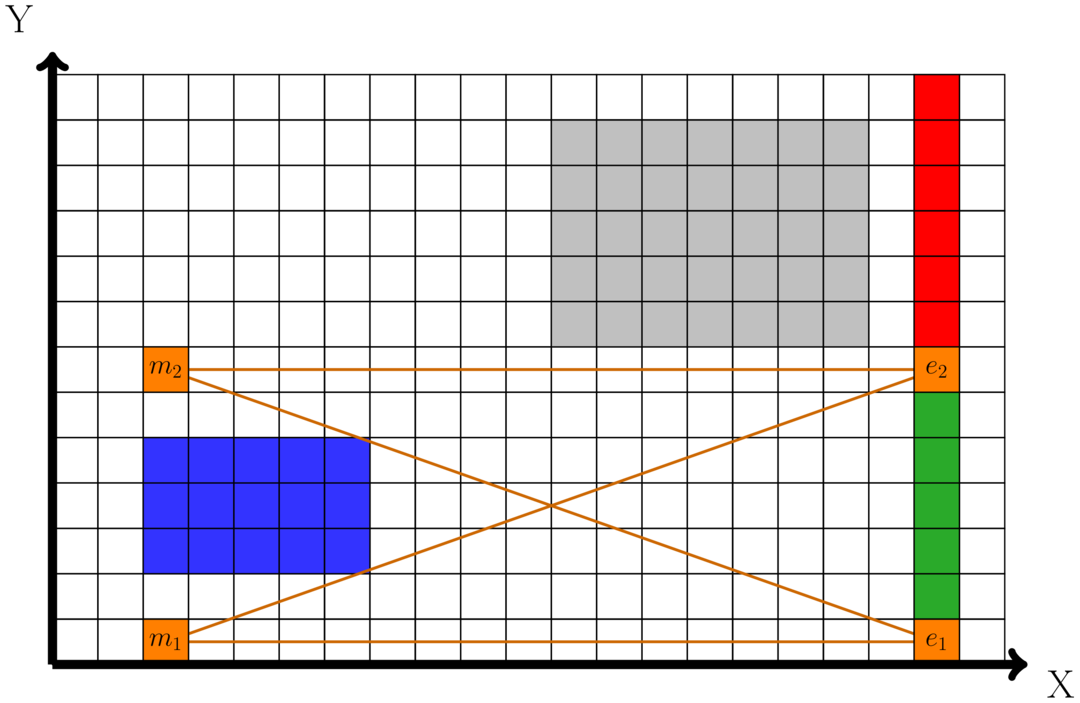}}
	\\
	\subfloat[The light-blue area displays the convex hull, the vertices are marked with orange cells. \label{hull:3}]{\includegraphics[scale=0.20]{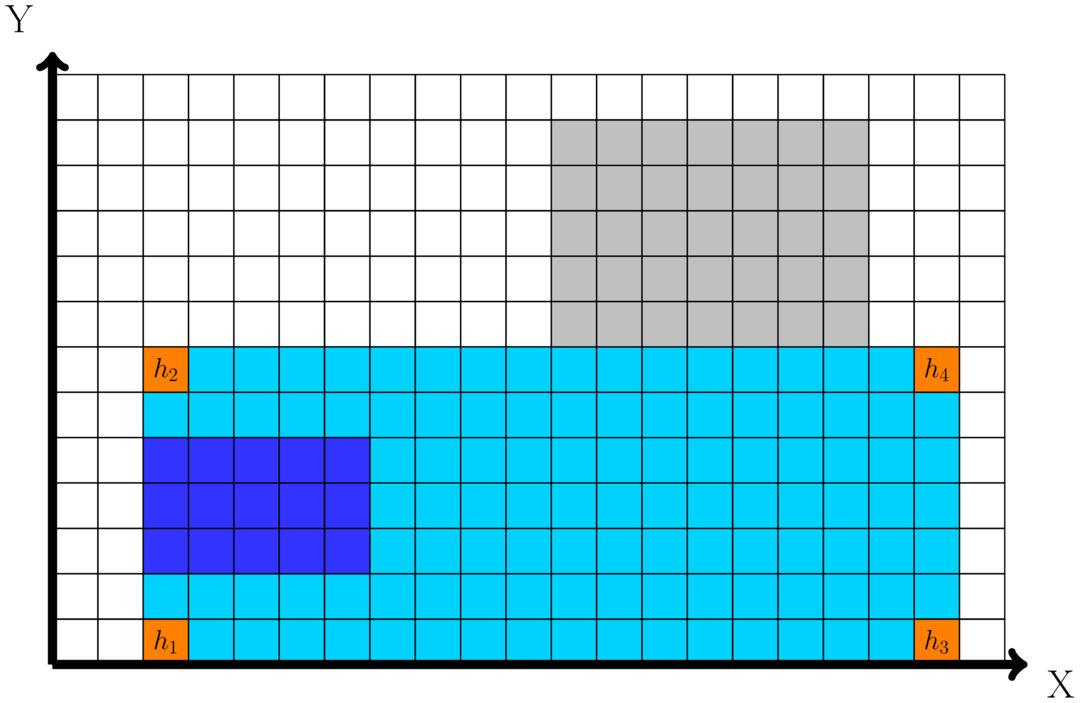}}
	\caption{Illustration of the steps in Algorithm~\ref{algsafe}.}
	\label{hull:all}
\end{figure}

The basic steps are shown in Figure~\ref{hull:all}, where the blue cells represent the \gls{ev} vehicle and the gray cells represent the inadmissible space according to \eqref{eq:bog}. The scenario is a straight road with direction of motion along the $x$-axis. First, valid cells in the \gls{ev} detection range are determined (Figure~\ref{hull:1}). Then, using Bresenham's algorithm, it is checked whether the cells on the detection range boundary allow straight connecting lines to the \gls{ev} without intersecting occupied cells. Second, the area of valid cells around the \gls{ev} is enlarged (Figure~\ref{hull:2}), resulting in a convex hull (Figure~\ref{hull:3}). The detailed approach is described in Algorithm~\ref{alg:1}. 
\begin{algorithm}
	\caption{Admissible Safe State Space search} \label{algsafe}
	\begin{algorithmic}[1]
		\renewcommand{\algorithmicrequire}{\textbf{Input:}}
		\renewcommand{\algorithmicensure}{\textbf{Output:}}
		\REQUIRE $\gls{xiev}_{h}$, $\mathcal{B}_{h}$
		\STATE Identify \gls{ev} center $\boldsymbol{c}^\text{EV}$ and select rear corners $c_{\text{r},1}$ and $c_{\text{r},2}$ (as in Figure~\ref{hull:1})
		\STATE Select admissible cells in the detection range of the \gls{ev} and store them in a matrix $\boldsymbol{C}_\text{range}$  (cyan column in Figure~\ref{hull:1})
		\FORALL{cell $\boldsymbol{c}$ in $\boldsymbol{C}_\text{range}$ } 
			\STATE{Verify free path from $\boldsymbol{c}$ to $c_{\text{r},1}$, and from $\boldsymbol{c}$ to $c_{\text{r},2}$} 
			\IF{free paths are confirmed} 
			    \STATE {Store $\boldsymbol{c}$ in matrix $\boldsymbol{C}_\text{free}$}
		    \ENDIF
		\ENDFOR
		\STATE Select edge cells $e_1$ and $e_2$ of $\boldsymbol{C}_\text{free}$ (as in Figure~\ref{hull:2})
		\STATE Define $m_1$ as $c_{\text{r},1}$ and $m_2$ as $c_{\text{r},2}$ 
		\STATE Move down $m_1$ as long as free path exists between $m_1$, $e_1$ and $m_1$, $e_2$
		\STATE Move up $m_2$ as long as free path exists between $m_2$, $e_1$ and $m_2$, $e_2$
		\STATE Select vertices of the convex hull $h_1$, $h_2$, $h_3$, $h_4$
		\ENSURE Convert vertex-representation into linear inequality representation \eqref{hull}
	\end{algorithmic} 
	\label{alg:1}
\end{algorithm}

Algorithm~\ref{algsafe} is used for each time step $h$ of the prediction horizon $N$. It results in a linear inequality description of a convex hull, i.e.,
\begin{IEEEeqnarray}{rl}
	\IEEEyesnumber \label{hull}
	\boldsymbol{A}^{\text{safe}}_{h} \gls{xiev}_{h} &\leq \boldsymbol{b}^{\text{safe}}_{h}.
\end{IEEEeqnarray}
\noindent
This safety constraint is now included in an MPC optimal control problem. Note that there is no guarantee that an admissible convex hull is found for each prediction step $h$ within the prediction horizon $N$. It is, however, assumed that a convex hull can always be found for $h=1$. If no admissible convex hull is found at prediction step $h$, the convex hull of step $h-1$ is used.

\subsection{Model Predictive Controller}
Given the convex hull representing the admissible \gls{ev} states, we can formulate a tractable \gls{SMPC} optimization problem, which is denoted as a standard \gls{MPC} optimization problem. The optimization problem is given by
\begin{IEEEeqnarray}{rl}
	\IEEEyesnumber  \label{eq:ocp}
	 \underset{\boldsymbol{U}}{\min} & \sum_{h=0}^{N-1} \left(\norm{\Delta \boldsymbol{\xi}^\text{EV}_{h}}^2_{\boldsymbol{Q}} + \norm{\boldsymbol{u}_{h}^\text{EV}}^2_{\boldsymbol{R}} \right) +\norm{\Delta \boldsymbol{\xi}^\text{EV}_{N}}^2_{\boldsymbol{S}}  \IEEEyessubnumber  \IEEEeqnarraynumspace\\	
	\text{s.t.}  & \quad \boldsymbol{\xi}_{h+1}^\text{EV} = f\left(\boldsymbol{\xi}_{h}^\text{EV},\boldsymbol{u}_{h}^\text{EV}\right), \quad h \in \mathbb{N}, \IEEEyessubnumber\\
	& \quad \gls{xitv}_{h+1} = \boldsymbol{A} \gls{xitv}_{h} + \boldsymbol{B}\bm{u}_{h}^\text{TV} + \boldsymbol{G} \bm{w}_{h}^\text{TV}, \quad  h \in \mathbb{N}, \IEEEyessubnumber\\
	& \quad \boldsymbol{u}_h^\text{TV} = \boldsymbol{K} \lr{\gls{xitv}_h - \gls{xitv}_{h,\text{ref}}}, \quad  h \in \mathbb{N},\IEEEyessubnumber\\
	& \quad \boldsymbol{u}_{h}^\text{EV} \in \mathcal{U}_{h}, \quad h=0,\ldots,N-1, \IEEEyessubnumber\\
	& \quad \boldsymbol{\xi}_{h}^\text{EV} \in \Xi_{h}, \quad h=1,\ldots,N, \IEEEyessubnumber\\
& \quad \boldsymbol{A}^{\text{safe}}_{h} \gls{xiev}_{h} \leq \boldsymbol{b}^{\text{safe}}_{h}, \quad h=1,\ldots,N, \IEEEyessubnumber \label{eq:ccinocp}
\end{IEEEeqnarray}
with the \gls{ev} input $\boldsymbol{U} = \left( \boldsymbol{u}_{0}, \ldots, \boldsymbol{u}_{N-1} \right)^\top$, $\norm{z}^2_{\boldsymbol{Z}} = z^\top \boldsymbol{Z} z$ and $\Delta \boldsymbol{\xi}_{h}^\text{EV} = \boldsymbol{\xi}_{h}^\text{EV} - \boldsymbol{\xi}^\text{EV}_{h,\text{ref}}$ with maneuver-dependent \gls{ev} reference 
$\boldsymbol{\xi}^\text{EV}_{h,\text{ref}}$, weighting matrices $\boldsymbol{Q}, \boldsymbol{S} \in \mathbb{R}^{4\times 4} $ and $\boldsymbol{R}~\in~\mathbb{R}^{2\times2}$, \gls{ev} prediction model $f\left(\boldsymbol{\xi}_{h}^\text{EV},\boldsymbol{u}_{h}^\text{EV}\right)$ according to \eqref{eq:systemEV_general}, and the input and state constraints $\mathcal{U}_{h}$ and $\Xi_{h}$. The \gls{tv} model for the prediction is according to \eqref{eq:systemTV}, \eqref{eq:uTV}.

In this optimization problem, the chance constraint \eqref{eq:chance} is reformulated into the tractable safety constraint \eqref{eq:ccinocp} according to \eqref{hull}.

%% file: chapters/results.tex
\section{Results}
\label{sec:results}

This \gls{SMPC} algorithm is now applied to an autonomous driving scenario. In the following, we first provide the general setup that has been used for the simulations. Then, we present the results of a first simulation, a highway scenario where an \gls{ev} overtakes two \glspl{tv}, to demonstrate the method. Eventually, we show the results of a second simulation with multiple \glspl{tv} to analyze the computational cost obtained with the proposed grid-based \gls{SMPC} method. 
\subsection{Simulation Setup}

The presented method has been implemented in MATLAB\textsuperscript{\textregistered} using the \acrshort{MPC} routine developed in \cite{GruenePannek2017} as a base implementation. Each vehicle has been modeled as a rectangle with length and width of \SI{6}{\metre} and \SI{2}{\metre}. Simulations are run with sampling time $\Delta t = \SI{0.2}{\second}$. 
The scenario is a two-lane highway with a lane width $l_\text{lane} =  \SI{3.5}{\metre}$. Cell dimensions are $\gls{cx} = \SI{0.5}{\metre}$ and $\gls{cy} = \SI{0.25}{\metre}$. 

We consider the \gls{ev} model
\begin{IEEEeqnarray}{rl}
	\IEEEyesnumber \label{eq:systemEV}
	\dot{x} &= v \cos (\psi + \alpha), \IEEEyessubnumber \\
	\dot{y} &= v \sin (\psi + \alpha), \IEEEyessubnumber \\
	\dot{\psi} &= \frac{v}{\gls{lr}} \sin(\alpha), \IEEEyessubnumber  \\
	\dot{v} &= a, \IEEEyessubnumber \\
	\alpha &= \arctan \left(\frac{\gls{lr}}{\gls{lr} + \gls{lf}} \tan \left(\gls{df}\right) \right) \IEEEyessubnumber 
\end{IEEEeqnarray}
\noindent
where $x$ and $y$ represent the longitudinal and lateral position of the vehicle \gls{cog}, $v$ and $a$ denote the longitudinal velocity and acceleration, $\alpha$ denotes the body slip angle, 
$\delta_\text{f}$ is the steering angle of the front wheels, \gls{lr} and \gls{lf} are the distances from the vehicle \gls{cog} to the rear and front axles, respectively. We denote with $\gls{xiev} = \left[ x,y,\psi, v \right]^\top$ and $\boldsymbol{u}^\text{EV} = \left[\gls{df},a \right]^\top$ the state and the control input of the \gls{ev}. Model \eqref{eq:systemEV} is discretized by the forward-Euler method with sampling time $\Delta t$.

The \gls{ev} employs two simple policies to decide on a reference lane $y^\text{EV}_{h,\text{ref}}$: 1) if the actual lane is occupied \SI{20}{\metre} in front of it, the \gls{ev} moves to the nearest free lane, otherwise it keeps its actual lane, 2) when it passes a \gls{tv} and the longitudinal distance between their \glspl{cog} is larger than \SI{15}{\meter}, the \gls{ev} will overtake the \gls{tv} by positioning itself in front of the \gls{tv}.

The \gls{ev} is subject to the following constraints $\boldsymbol{u}^\text{EV} \in \mathcal{U}$,
\begin{IEEEeqnarray}{rl}
	\IEEEyesnumber \label{eq:evconstr}
	y~ &\in \SIrange{1}{6}{\meter}, \IEEEyessubnumber\\
	\delta_{\text{f}}~ &\in \SIrange{-3}{3}{\deg}, \IEEEyessubnumber\\ 
	a~ &\in \SIrange{-5}{5}{\meter / \second^2}, \IEEEyessubnumber
\end{IEEEeqnarray}
\noindent in addition to the safety inequality constraints \eqref{hull}.

We consider a time-discrete point-mass \gls{tv} prediction model for \eqref{eq:systemTV} with 
\begin{IEEEeqnarray}{c}
    \label{eq:tvmatrices}
    \resizebox{0.4\textwidth}{!}{%
	$\boldsymbol{A} = \begin{bmatrix} 1 & \Delta t & 0 & 0\\ 0 & 1 & 0 & 0\\ 0 & 0 & 1 & \Delta t \\ 0 & 0 & 0 & 1 \end{bmatrix}, ~~
	\boldsymbol{B} = \begin{bmatrix} 0.5 \lr{\Delta t} ^2 & 0\\ \Delta t & 0\\ 0  & 0.5 \lr{\Delta t} ^2 \\ 0 & \Delta t \end{bmatrix}.$
	}
\end{IEEEeqnarray}
The selected \gls{tv} controller matrix values are $\left[k_{12},k_{21},k_{22}\right] = \left[-1, -0.8, -2.2\right]$. We assume Gaussian noise $\boldsymbol{w}_k^\text{TV}~\sim~\mathcal{N}(\boldsymbol{0},\boldsymbol{\Sigma}_w)$ with covariance matrix $\boldsymbol{\Sigma}_w = \text{diag}\lr{1,1,1,1}$ and disturbance matrix $\boldsymbol{G} = \text{diag}\lr{0.05, 0.067, 0.013, 0.03}$. These choices for the TVs are similar to \cite{BruedigamEtalWollherr2018b}. 

Here, only the two most likely \gls{tv} maneuvers are considered: a \gls{lk} and a \gls{lc} maneuver with constant longitudinal velocity where each of them is weighted with a probability value for the respective maneuver being executed, as shown in Sec. \ref{sec:scenarioext}. Note that more maneuvers could be considered. Here, we randomly assign a probability in the range of \SIrange{0.8}{1}{} to one of the predicted maneuvers. The second maneuver is given a probability such that the sum equals one.

The \gls{SMPC} has a prediction horizon $N=20$, weighting matrices $\boldsymbol{Q} = \text{diag}\lr{0,2,0.5,0.1}$ and $\boldsymbol{R} = \text{diag}\lr{0.1,1}$, and a probability threshold $\gls{pth} = 0.15$. 

Algorithm \ref{algsafe} is used to find a convex hull at each time step~$h$ of the prediction horizon $N$. If at a generic time step $h_t$ a convex hull is not found, we consider the one calculated at the preceding time step $h_{t}-1$. Therefore, Algorithm \ref{algsafe} is based on the assumption that a convex hull can always be found at step $h=0$.\\

\subsection{Overtaking Scenario} 
\label{sec:over_scen}

This scenario consists of a straight two-lane road with two \glspl{tv}. The center of the right lane is set to \SI{1.75}{\meter}, the left lane to \SI{5.25}{\meter}. The \gls{ev} is positioned on the left lane with initial state $\gls{xiev} = \left[10, \gls{lref}, 0, 26 \right]$, where \gls{lref} = \SI{5.25}{\meter}. The two \glspl{tv} start with the initial states $\gls{xitv1} = \left[40, 27, 5.25, 0 \right]$, and $\gls{xitv2} = \left[90, 27, 1.75, 0 \right]$. \gls{tv1} is positioned on the left lane, \gls{tv2} on the right one. A probability value 0.8 is assigned to the \gls{lk} maneuver and 0.2 to the \gls{lc} maneuver for both \glspl{tv}. In the simulation the \glspl{tv} follow the maneuver with the higher probability. Therefore, both \glspl{tv} will actually proceed along their lane. 

\begin{figure}[!htbp]
	\centering
	\subfloat[The \gls{ev} moving to the free lane to avoid collision with \gls{tv1}. \label{shot:1}]{\includegraphics[scale=0.38,trim = 26 0 0 0]{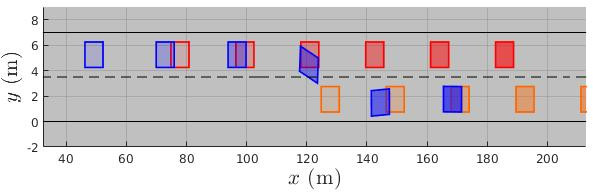}} 
	\\
	\subfloat[\gls{ev} overtaking \gls{tv1}. \label{shot:2}]{\includegraphics[scale=0.38,trim = 26 0 0 0]{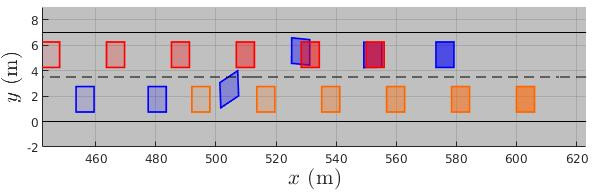}}
	\\
	\subfloat[\gls{ev} overtaking \gls{tv2}. \label{shot:3}]{\includegraphics[scale=0.38,trim = 26 0 0 0]{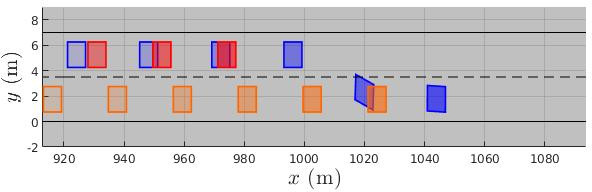}}
	\\
	\subfloat[Visualization of the vehicle motion for the first screenshot.
	\label{shot:traj1}]{\includegraphics[scale=0.38,trim = 26 0 0 0]{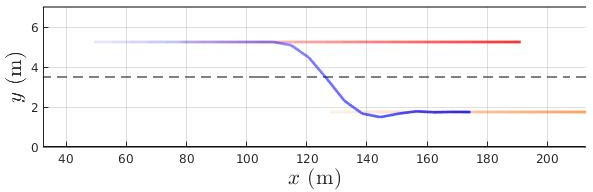}}
	\\
	\subfloat[Visualization of the vehicle motion for the second screenshot.
	\label{shot:traj2}]{\includegraphics[scale=0.38,trim = 26 0 0 0]{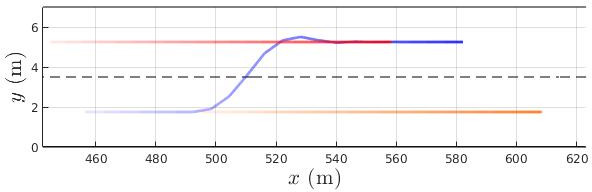}}
	\\
	\subfloat[Visualization of the vehicle motion for the third screenshot.
	\label{shot:traj3}]{\includegraphics[scale=0.38,trim = 26 0 0 0]{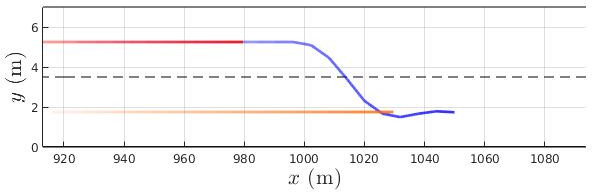}}
	\\
	\caption{Screenshot of the simulation during the overtaking maneuvers. Blue boxes represent the \gls{ev}, red and orange boxes represent \gls{tv1} and \gls{tv2}, respectively. Fading boxes show past states.}
	\label{shot:all}
\end{figure}

\begin{figure}[!htbp]
	\centering
	\includegraphics[scale=0.45]{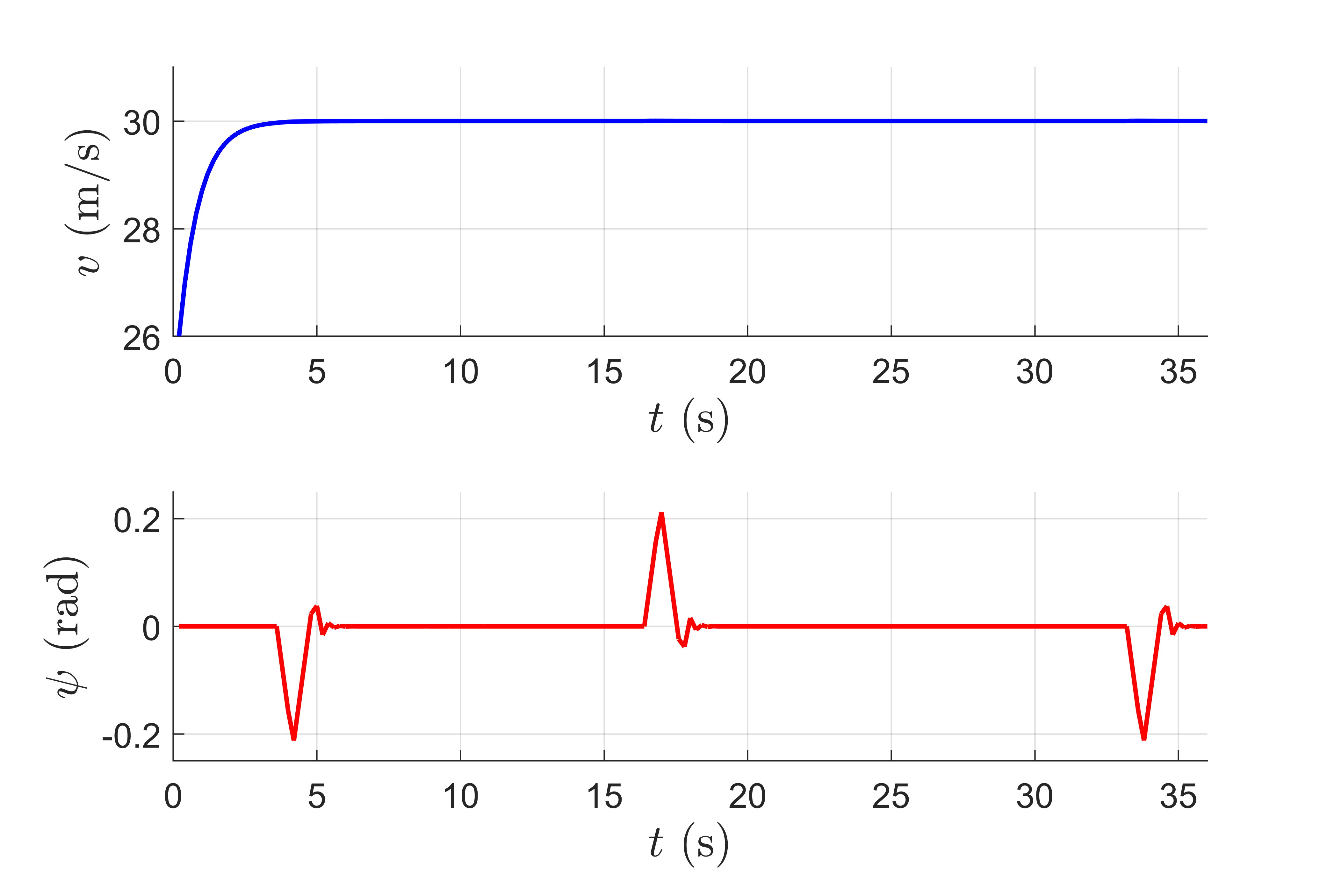}
	\caption{\gls{ev} longitudinal velocity and steering.}
	\label{evvel}
\end{figure}

\begin{figure}[!htbp]
	\centering
	\subfloat[Distance between \gls{ev} and \gls{tv1}. \label{posdiff:1}]{\includegraphics[scale=0.45]{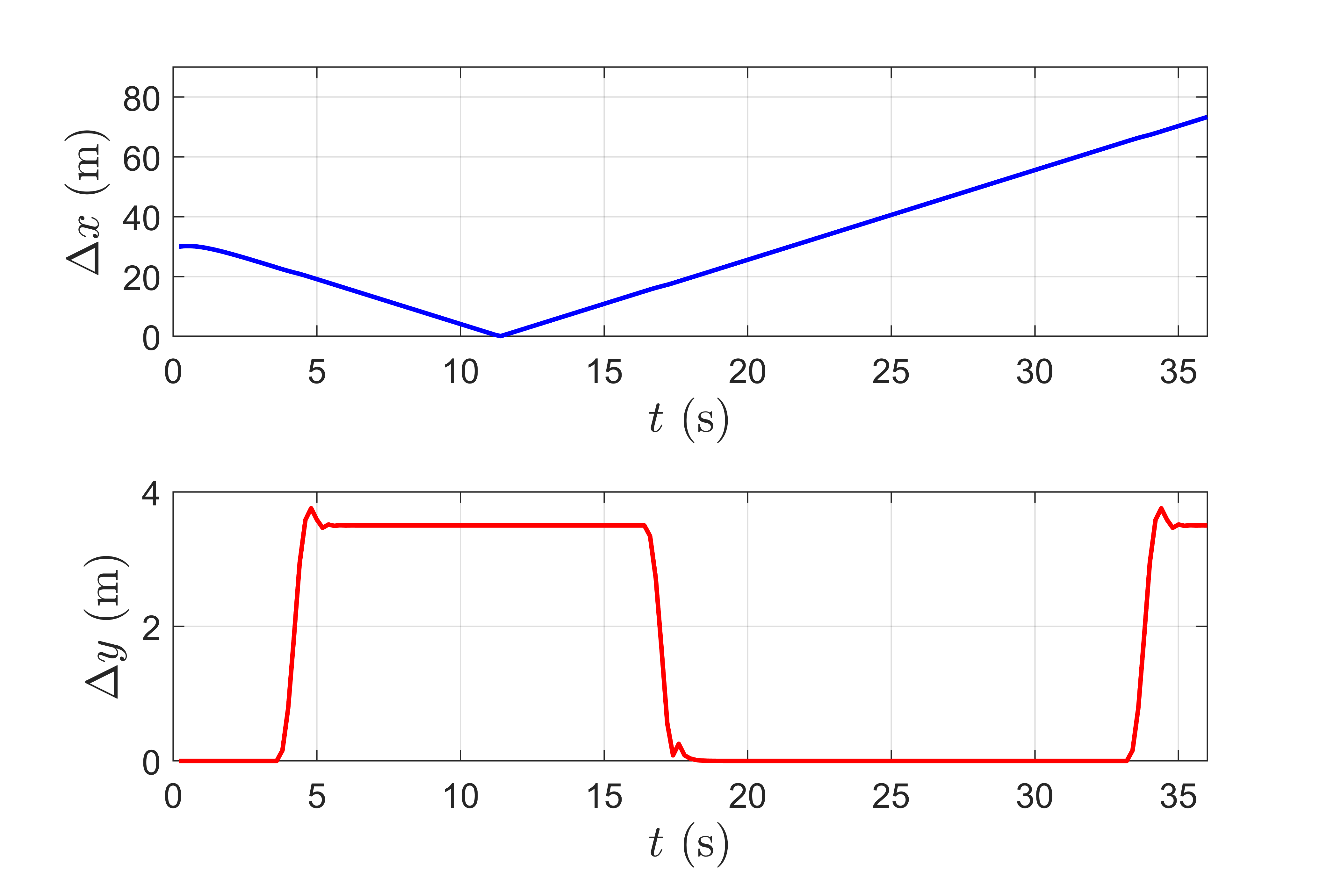}}
	\\
	\subfloat[Distance between \gls{ev} and \gls{tv2}. \label{posdiff:2}]{\includegraphics[scale=.45]{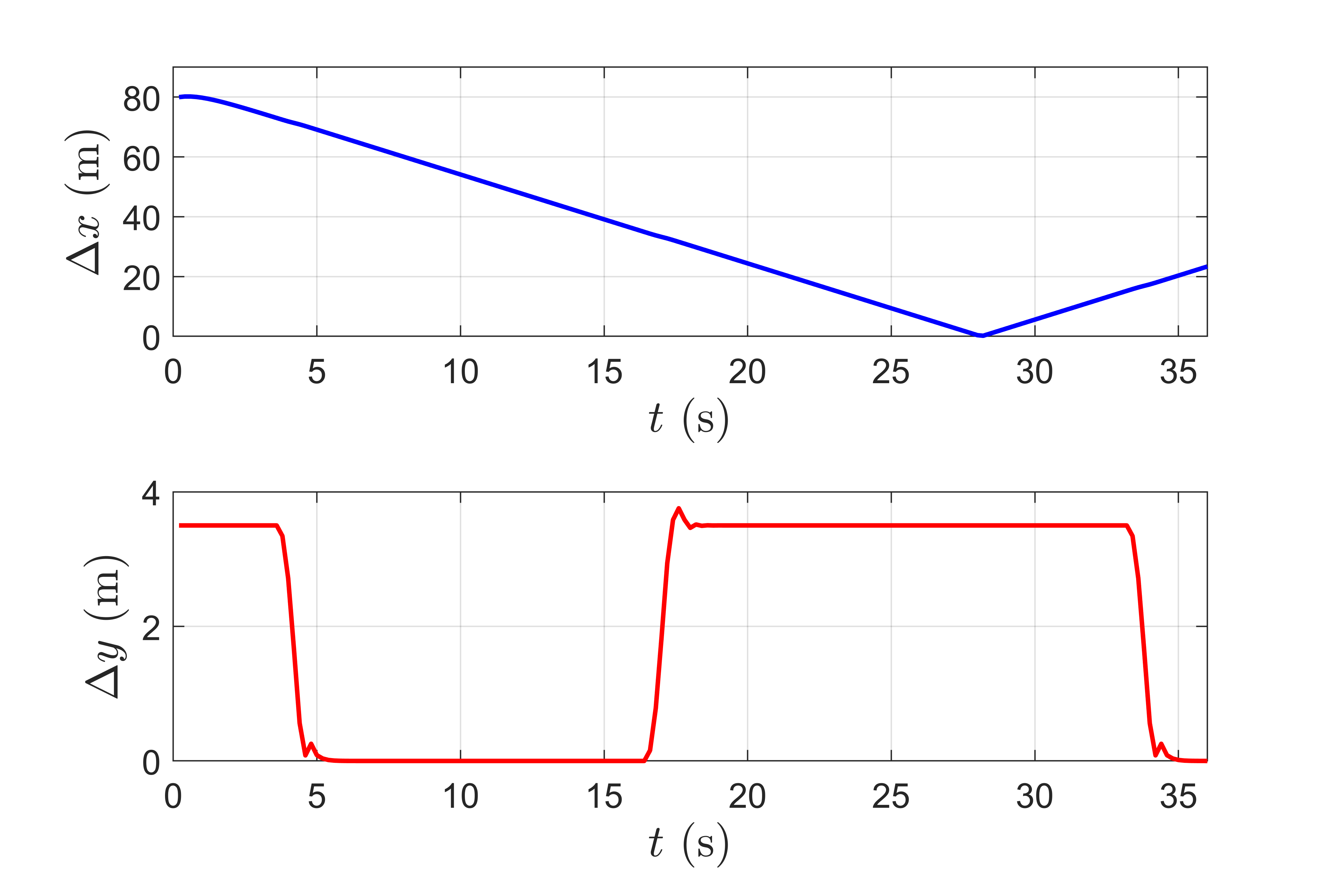}}
	\caption{Longitudinal and lateral distance between \gls{ev} and \glspl{tv} in absolute value terms.}
	\label{posdiff:all}
\end{figure}

Figures \ref{shot:all}, \ref{evvel}, and \ref{posdiff:all} show the simulation results. Figure~\ref{shot:all} illustrates the vehicle motion, Figure~\ref{evvel} displays the \gls{ev} velocity and steering angle, while Figure~\ref{posdiff:all} shows the distance between the \gls{ev} and the two \glspl{tv}. At the beginning the \gls{ev} accelerates to reach its reference velocity of \SI{30}{\metre/\second}, as shown in the first plot of Figure~\ref{evvel}. When the lane is occupied by \gls{tv1} \SI{20}{\metre} in front of the \gls{ev}, the \gls{ev} starts a \gls{lc} maneuver, moving to the right lane, as shown in Figure~\ref{shot:1}. In Figure~\ref{posdiff:1} it can be seen that once the maneuver is completed, the longitudinal distance between \gls{ev} and \gls{tv1} is approximately \SI{16}{\metre}. As soon as it passes \gls{tv1} and the distance between their \glspl{cog} is larger than \SI{15}{\metre}, the reference lane for the \gls{ev} changes and the \gls{ev} starts to move towards the left lane to finish overtaking \gls{tv1}. Figure~\ref{shot:2} shows a screenshot of this phase. When the \gls{ev} has completed the maneuver, i.e., when it lies completely in the left lane, the longitudinal distance with respect to \gls{tv1} is about \SI{17}{\metre}. Once it reaches and passes \gls{tv2}, the \gls{ev} performs a new \gls{lc} maneuver by moving to the right lane again. Once the \gls{ev} lies on the right lane, the relative distance between the \glspl{cog} of the two vehicles is approximately \SI{21}{\metre}. This last phase is represented in Figure~\ref{shot:3}.

For the presented scenario, and with the given policies, we can see that there is no deceleration by the \gls{ev} while performing \gls{lc} maneuvers, but only changes in the steering angle $\psi$. Collisions are avoided. As the simulated (not the predicted) \gls{tv} motion is deterministic in this scenario, one simulation is sufficient for the proposed method. For the presented grid-based \gls{SMPC} method, additional simulations with identical initialization result in the same behavior, in contrast to \gls{SMPC} methods based on sampling.

Given this setup and the policy to compute the reference lane for the \gls{ev}, the MATLAB\textsuperscript{\textregistered} solver \textit{fmincon} always finds a feasible solution. Therefore, bounds on control signal and space constraints are respected. However, it is important to mention that these results are obtained with the choice $\gls{pth} = 0.15$, and no further research has been conducted on different values. The effect of varying risk parameters is studied in other \gls{SMPC} works, e.g., \cite{CarvalhoEtalBorrelli2014, BruedigamEtalWollherr2018b}.

\subsection{Computational cost evaluation} 
To evaluate the computational cost of the algorithm, we use the following more complex setup. The general setup is the one shown in Section~\ref{sec:over_scen}. For each simulation, the \gls{ev} is randomly positioned on one of the two lanes and a random reference lane is assigned. The same applies to each \gls{tv}. The first \gls{tv} is positioned in front of the \gls{ev} with a longitudinal distance of \SI{40}{m}. If more \glspl{tv} are simulated, these are positioned every \SI{50}{m}. For each \gls{tv}, one probability value is sampled in the range \SIrange{0.8}{1}{} and assigned randomly to one of the two maneuvers, LC or LK, and the second value is assigned to the other maneuver such that the sum of the probabilities equals one. The actual behavior of a \gls{tv} follows the maneuver with higher probability. We simulated three scenarios with one, two, and three \glspl{tv}, and each of them has been run 10 times on a standard desktop computer with an Intel i5 processor ($\SI{3.3}{GHz}$).\\
\begin{table}[!htbp]
	\caption{Mean and standard-deviation per algorithm iteration for scenarios with a different number of TVs .}
	\begin{center}
	\renewcommand{\arraystretch}{1.6}
	\label{tab:tab1}
	\begin{tabular}{l|c|c}
		\hline
		\# of \glspl{tv} & $\mu$ \SI{}{(\second)} & $\sigma$ \SI{}{(\second)} \\
		\hline 
		1 \gls{tv} & 0.45 & 0.48 \\
		\hline 
		2 \glspl{tv} & 0.44 & 0.43 \\
		\hline 
		3 \glspl{tv} & 0.42 & 0.36 \\
		\hline 
	\end{tabular}
	\end{center}
\end{table}
\noindent
Table \ref{tab:tab1} shows the mean $\mu$ and the standard deviation $\sigma$ of the algorithm computation time per iteration. By increasing the number of \glspl{tv} in the \gls{ev} detection range for the scenario, the computation cost mean $\mu$ remains almost constant, while the standard deviation $\sigma$ shows larger variations. 
The computational cost of the algorithm is mainly due to the complexity of the nonlinear \gls{ev} model \eqref{eq:systemEV}, and not dependent on the number of \glspl{tv} in the scenario. The computational effort generating the \gls{pog} and calculating the convex hull is comparatively small, as this is done prior to solving the optimization problem. This is a major advantage over for example \cite{BruedigamEtalWollherr2018b}, where the computation time increases significantly with an increasing number of \glspl{tv}.

\subsection{Discussion}

In Section \ref{sec:methodc} we introduced the \textit{probability threshold} parameter \gls{pth}, which allows to transform the \gls{pog} into a \gls{bog}, in order to obtain a deterministic approximation of the chance constraints \eqref{eq:chance}. This parameter is a trade-off between conservatism and risk. By setting a low value for \gls{pth}, a high number of cells will be considered occupied. At a certain step, the road can seem fully occupied, resulting in a conservative maneuver for the \gls{ev}. On the other hand, a high value of \gls{pth} will reduce the number of occupied cells considered by the algorithm and, at certain step of the prediction horizon, the road can seem free. This can yield a more aggressive controller and potentially a collision between the vehicles.

Therefore, the parameter \gls{pth} has to be chosen by evaluating a trade-off between conservatism and risk for the planning algorithm, depending on the kind of distribution one decides to adopt. It can also be beneficial to apply a time-varying probability threshold, adapting to different situations. In general, selecting a suitable threshold is challenging, similar to choosing a risk parameter in other \gls{SMPC} approaches. Note that the risk parameter and the considered probability values do not perfectly represent true probabilities.

In the simulations a nonlinear \gls{ev} model is used to increase the accuracy of the \gls{ev} predictions, whereas a simpler, linear \gls{tv} model is used in combination with noise, as the \gls{tv} behavior is subject to uncertainty. However, applying a linearized \gls{ev} prediction model allows to solve a QP problem, given the safety constraint \eqref{hull}, a quadratic cost function, as well as linear input and state constraints. This is highly beneficial when a fast algorithm is needed that still considers stochastic behavior of surrounding vehicles.

%% file: chapters/discussion.tex

%% file: chapters/conclusion.tex
\section{Conclusion}
\label{sec:conclusion}

We presented a novel and simple approach to apply \gls{SMPC} for trajectory planning in uncertain environments, by using a probabilistic grid. This allows efficiently planning ego vehicle trajectories, while considering stochastic behavior of surrounding objects. The proposed method scales well with an increasing number of objects considered, here shown for three vehicles, and can handle arbitrary probability distributions of future object motion. It is still of interest to obtain simulation results for more complex scenarios and target vehicle maneuvers, as well as combine the proposed approach with occupancy grids.

While we applied the proposed method to autonomous vehicles, other applications are possible. It is especially interesting to apply the grid-based \gls{SMPC} approach to three-dimensional applications, e.g., trajectory planning for robots.